\documentclass{article}

\usepackage{hyperref}
\usepackage{graphicx}
\usepackage{fullpage}
\usepackage{authblk}

\begin{document}

\title{Characterisation and testing of a prototype $6 \times 6$ cm$^2$ Argonne MCP-PMT}

\author[1]{G.~A.~Cowan\footnote{g.cowan@ed.ac.uk}}
\author[1]{F.~Muheim}
\author[1]{M.~Needham}
\author[1]{S.~Gambetta}
\author[1]{S.~Eisenhardt}
\author[1]{N.~McBlane}
\affil[1]{School of Physics and Astronomy,
University of Edinburgh,
Edinburgh, UK}
\author[2]{M.~Malek}
\affil[2]{Department of Physics and Astronomy,
University of Sheffield,
Sheffield, UK}
\maketitle

\begin{abstract}
The Argonne micro-channel plate photomultiplier tube (MCP-PMT)
is an offshoot of the Large Area Pico-second Photo Detector (LAPPD) project,
wherein \mbox{6 $\times$ 6 cm$^2$} sized detectors are made at Argonne National Laboratory.
Measurements of the properties of these detectors, including gain, time and spatial resolution, dark count rates, cross-talk and sensitivity to magnetic fields are reported. In addition, possible applications of these devices in
future neutrino and collider physics experiments are discussed.
\end{abstract}

\section{Introduction}

\begin{figure}[h]
\centering{
\includegraphics[scale=0.075]{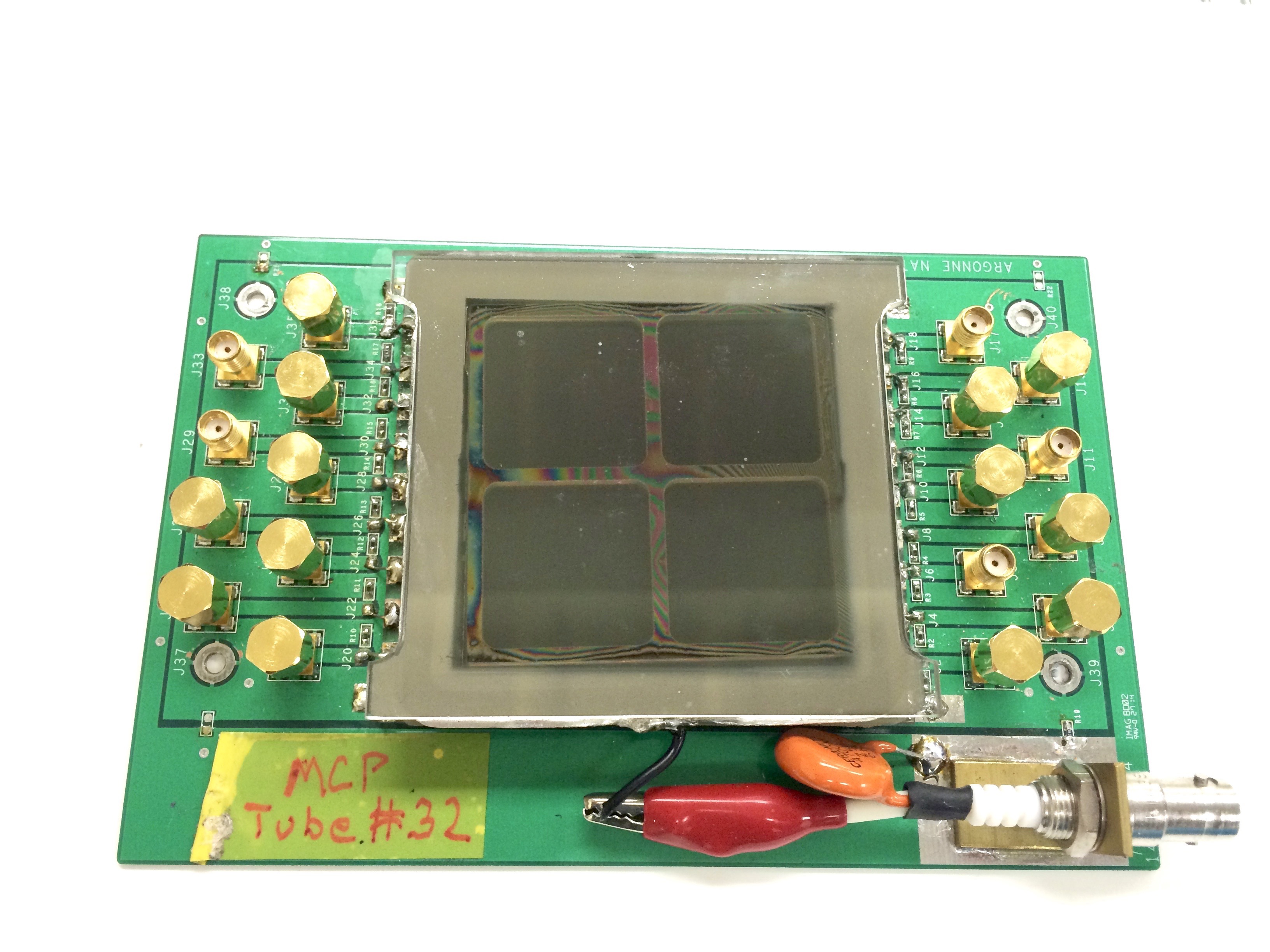}}
\caption{The prototype ANL MCP device. The active area can be seen 
in the centre of the device, with the cross-shaped spacer visible. The device
has nine anode strip lines (silk screened) at the base, which are read out 
on both ends using SMA connectors. Channels that are not read out
during a particular study are terminated using $50\,\Omega$ resistors.}
\label{fig:mcp}
\end{figure}

Micro-channel plate (MCP) detectors are sensitive to single photons, provide
picosecond-level time resolution, sub-mm level position resolution,
high gain ($10^6-10^7$) and low noise. 
The Large Area Picosecond PhotoDetector (LAPPD)
project~\cite{LAPPD} aimed to make the next generation of large area 
photodetectors using commercially viable techniques.
The Argonne MCP-based photodetector
has developed from the LAPPD project, wherein Argonne National Laboratory (ANL)
are producing {6 cm $\times$ 6 cm} devices.
The MCP borosilicate glass-substrates
are produced by Incom Inc. and are functionalised using  Atomic Layer Deposition
(ALD)~\cite{Mane:2012oxa},
which gives the substrates the correct resistive and secondary election emission
properties in order to function as an MCP.  Ref.~\cite{Wang:2016xnu}
describes a recent characterisation of these devices by ANL. They have sent
out prototype devices to early adopters in the high energy
physics community in order for them to be further evaluated and
potential applications developed. These proceedings describe initial tests
performed by the authors using an early prototype (``Tube 32"), kindly loaned to us
by ANL (Figure~\ref{fig:mcp}).

\section{Potential applications}

\begin{figure}[h]
\centering{
\includegraphics[scale=0.4]{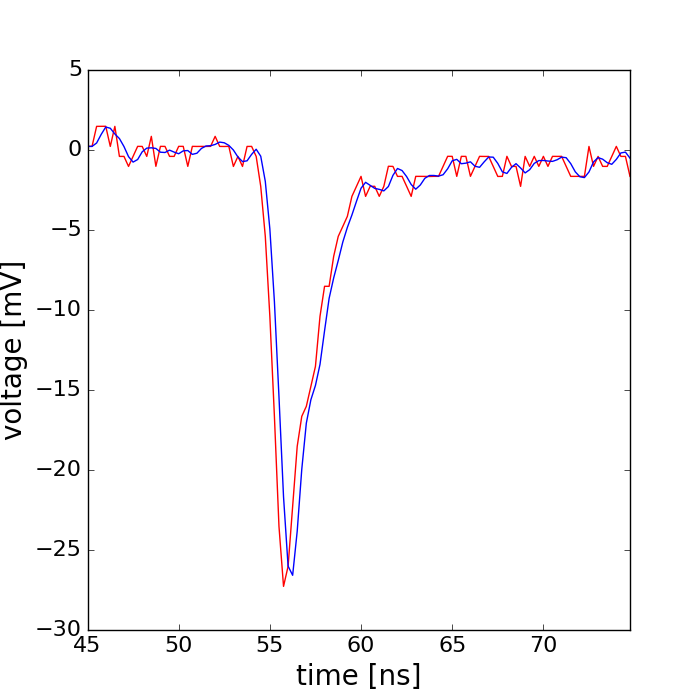}
\includegraphics[scale=0.4]{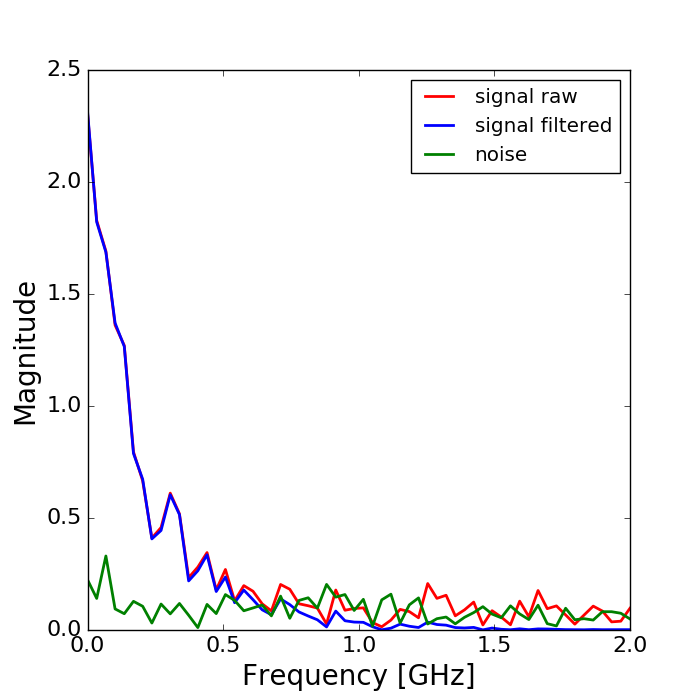}
}
\caption{Typical pulse shape with both the raw (red) and filtered (blue) signal (left). Fourier transform of the signal and background components (right).}
\label{fig:pulse}
\end{figure}

The ANL-MCPs allow fast timing, improved spatial resolution and have the potential
to tile large surface areas at reasonable cost.
This can lead to improved designs for future water Cherenkov~\cite{Anghel:2015xxt,Andreopoulos:2016rqc}
or liquid Argon~\cite{DUNE} neutrino experiments.
Simulations have shown~\cite{Anghel:2013zxa} that measuring photon arrival space-time
point with a resolution of 1 cm and 100 ps gives track and vertex reconstruction approaching
a few cm and better neutrino energy resolution.
Therefore, for small near or intermediate neutrino detectors, the use of MCPs maximises
the fiducial volume of the detector as the improved resolution allows the reconstruction
of neutrino interaction vertices closer to the walls of the detector.
There is also the potential for better discrimination between dark noise and photons from neutron capture.
Initial studies have already been performed~\cite{Dharmapalan:2016fai} to 
investigate how the ANL-MCP devices
behave in cryogenic temperatures suitable for liquid Argon applications.

The LHCb experiment~\cite{Alves:2008zz} is a dedicated heavy flavour physics
experiment at the CERN LHC. The collaboration is currently discussing possible
upgrades to the detector for the Run-4 and Run-5 periods of LHC operation
(from 2025 onwards). During this period, the luminosity of the LHC at the LHCb
interaction point will be approximately $10^{34}$ cm$^{-2}$s$^{-1}$ (roughly a 
factor twenty larger than the current luminosity). In this
high luminosity period, the occupancy in the central region of the first LHCb
RICH detector will be close to 100\%.
The RICH-1 intrinsic timing resolution is approximately 5 ps~\cite{private}, raising
the possibility to time the individual tracks associated with each proton-proton
collision and consequently reducing the effective occupancy of the detector.
LAPPDs or ANL-MCPs could provide the timing resolution required 
to associate photons with individual tracks passing through the detector.
Given the current dimension of RICH-1, one hundred $6 \times 6$ cm$^2$ ANL-MCPs,
or twelve $20 \times 20$ cm$^2$ LAPPDs, would be needed to tile the RICH-1 central region.

\section{Test facilities}

The Edinburgh photodetector test facilities consist of a 1 m$^3$
dark box where the device under test is mounted. It is 
illuminated using a pulsed 470 nm LED system, which
provides 10 ns wide pulses and can be operated in single or multi-photon mode. The
LED system is driven using a pulse generator. The light is
fed into the dark box via an optical fibre, which has an emission angle of 12$^\circ$
and is not focussed. Commercial NIM modules are used for the trigger system and two
different oscilloscopes are available for data acquisition (Tektronix 
6 GHz, 25 GS/s; LeCroy WavePro 960 1 GHz, 4 GS/s quad).
Custom LabView code is used to control the DAQ and offline processing is performed
using custom Python software~\cite{g_a_cowan_2016_160400}.

\section{Gain characterisation}

Figure~\ref{fig:pulse} (left) shows a typical MCP pulse when illuminated with the LED.
The MCP has a pulse width of approximately 5 ns @10\% of the
pulse maximum and a rise time of  $\sim0.5$~ns.
Ref.~\cite{Wang:2016xnu} reports a
transit time spread of 0.06 ns for single photons. 
As can be seen in Figure~\ref{fig:pulse} (right) a Butterworth low bandpass filter is
applied to remove noise above 700 MHz.

\begin{figure}[t!]
\centering{
\includegraphics[scale=0.45]{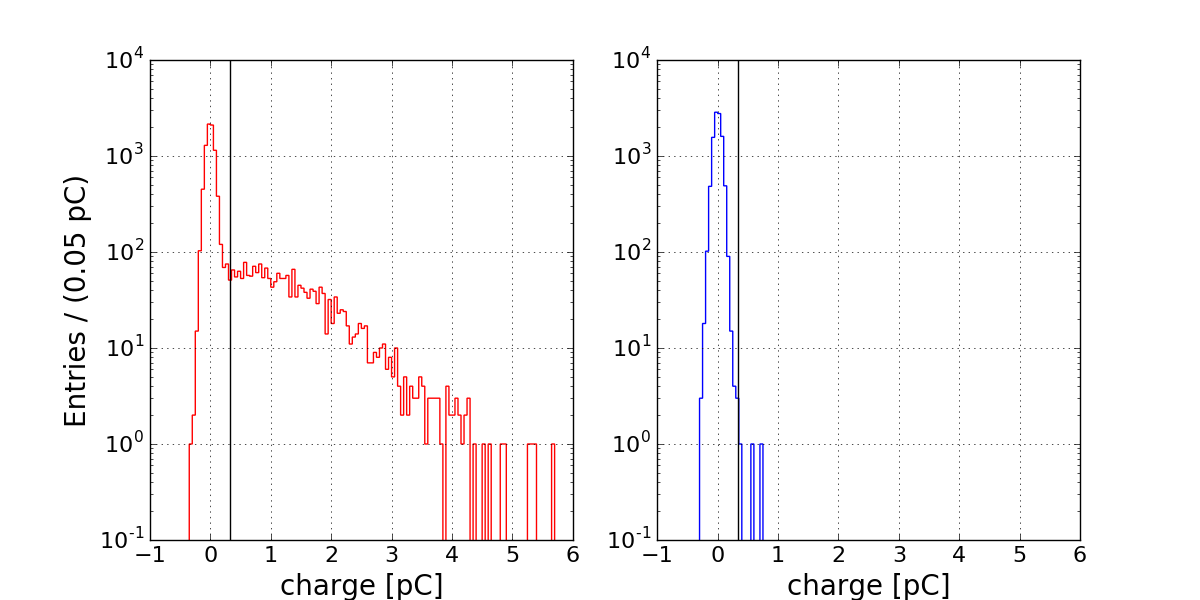}
}
\caption{Charge spectrum for channel J07 with LED on (left) and off (right). The vertical line shows the position
of the $5\sigma$ threshold. }
\label{fig:charge}
\end{figure}

\begin{figure}[t]
\centering{
\includegraphics[scale=0.45]{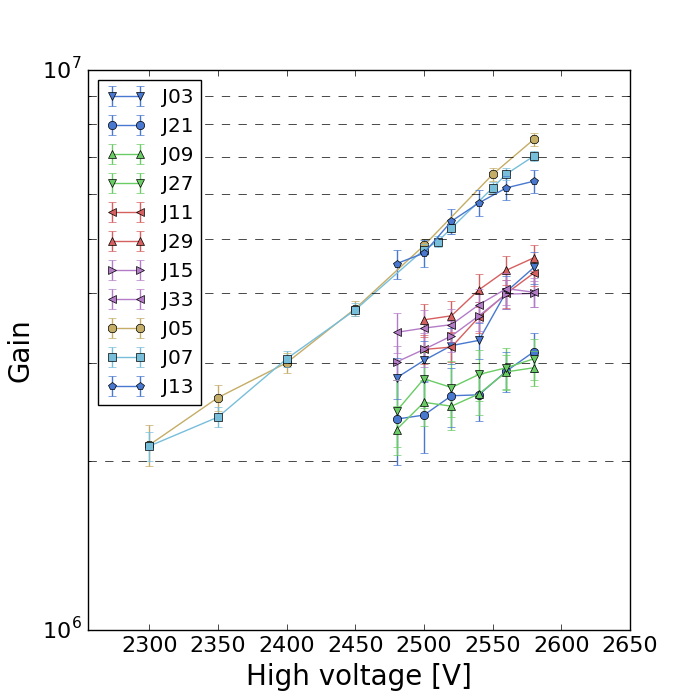}}
\caption{Gain as a function of high voltage for different channels on the device, 
labelled as JXY.
Channels with the same colour correspond to opposite ends of the same anode.
Channels J23 and J25 (opposite J05 and J07) were
not operational for this study.}
\label{fig:gain}
\end{figure}

From each event, the pulse shape is integrated to obtain the collected charge. These are 
used to build a distribution of collected charge for each run, examples of which
are shown in Figure~\ref{fig:charge} (a) and (b) for runs with the LED on and off, respectively.
The gain is estimated as the mean charge of events above a $5\sigma$ threshold,
which is determined from the pedestal collected when the MCP is not illuminated.
The separation between pedestal and single photon peak
varies between channels. This variation is visible in Figure~\ref{fig:gain}
which shows the gain as a function of the applied high voltage for most
of the channels on the device. Gains of $10^6-10^7$ are visible, but there
is some variation in gain between channels.
Below $\sim2450$ V the single photon peak signal is small.

\section{Gain as a function of magnetic field}

\begin{figure}[t]
\centering{
\includegraphics[scale=0.07]{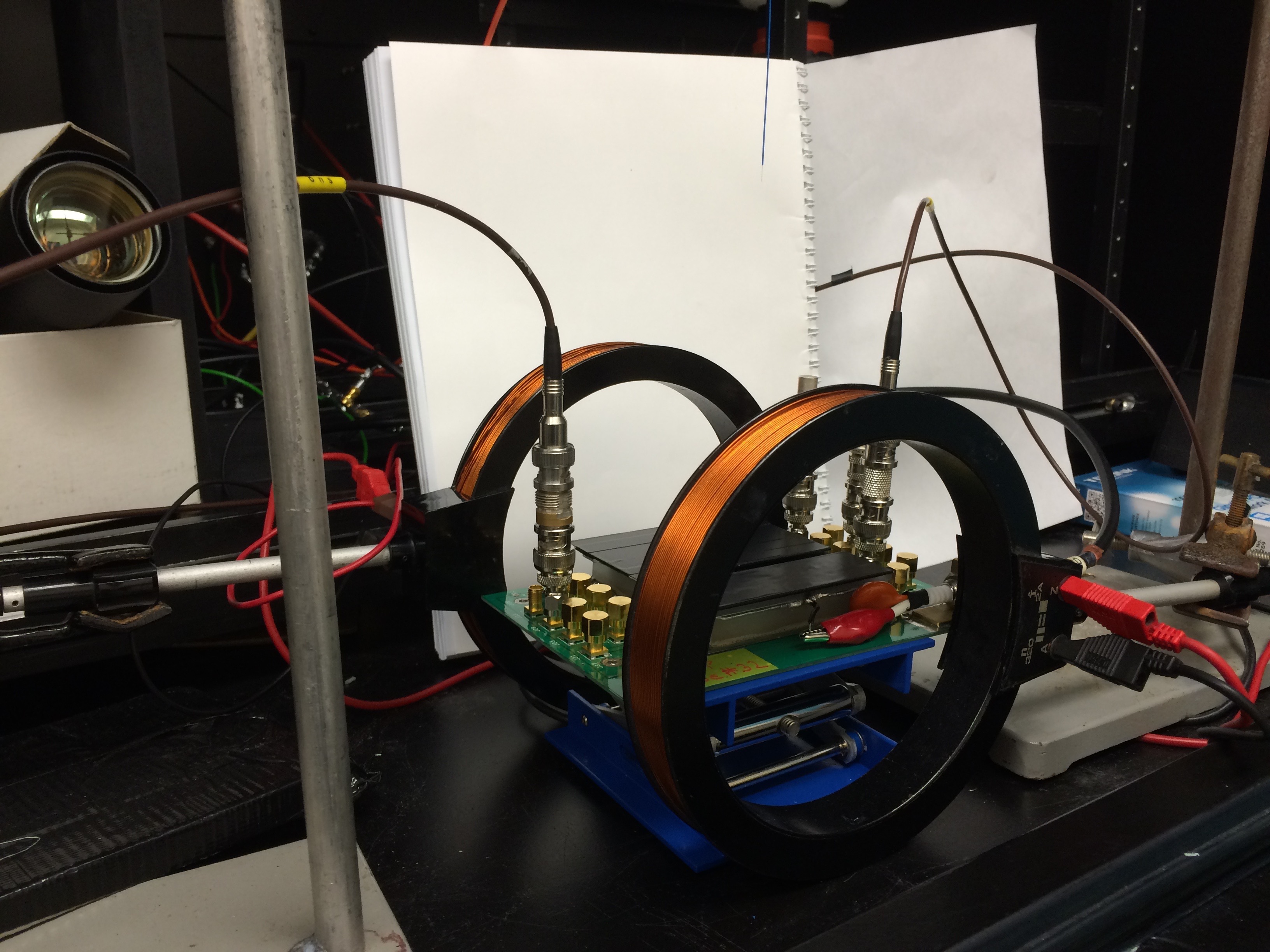}}
\caption{The MCP mounted in the dark box with the Helmholtz coils on either side in the transverse configuration.
The optical fibre is visible at the top of the picture.}
\label{fig:mcp_coils}
\end{figure}
\begin{figure}[h!]
\centering{
\includegraphics[scale=0.45]{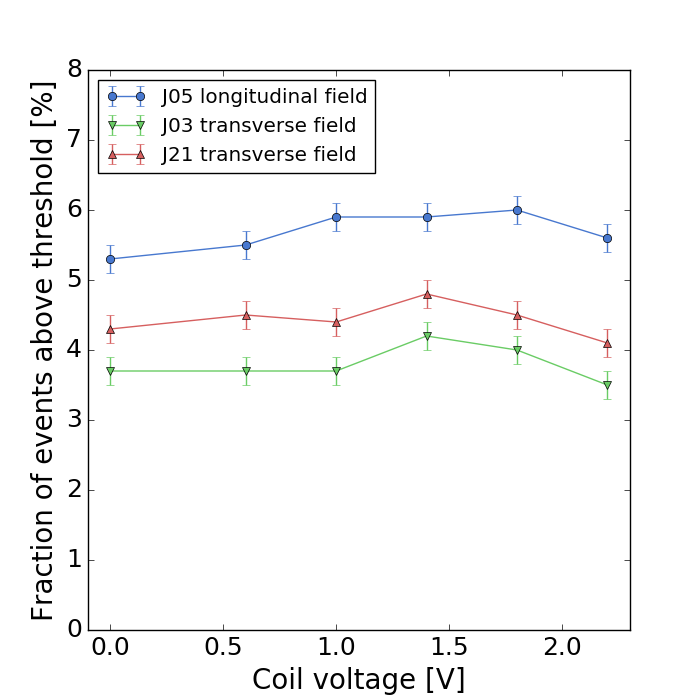}}
\caption{Variation of the fraction of events above threshold as a function of the voltage
applied to the Helmholtz coils (proxy for the magnetic field strength).}
\label{fig:gain_mag}
\end{figure}

As the MCPs could potentially be used in future particle physics experiments, it
is important to confirm that they are insensitive to magnetic fields. For example, 
if used in the LHCb RICH detector they will be subjected to an ${\cal O}(10)$ G
fringe field from the LHCb dipole magnet.
Using a pair of small Helmholtz coils we generated fields up to 30 G
in both a transverse (field perpendicular to the MCP capillaries, Figure~\ref{fig:mcp_coils})
and longitudinal configuration. 
Figure~\ref{fig:gain_mag} shows the fraction of events above the $5\sigma$ 
threshold as a function of the coil voltage (corresponding to a field of between 0
and 30 G). This clearly shows the insensitivity of the device to small magnetic fields.

\section{Gain as a function of temperature}

\begin{figure}[h!]
\centering{
\includegraphics[scale=0.45]{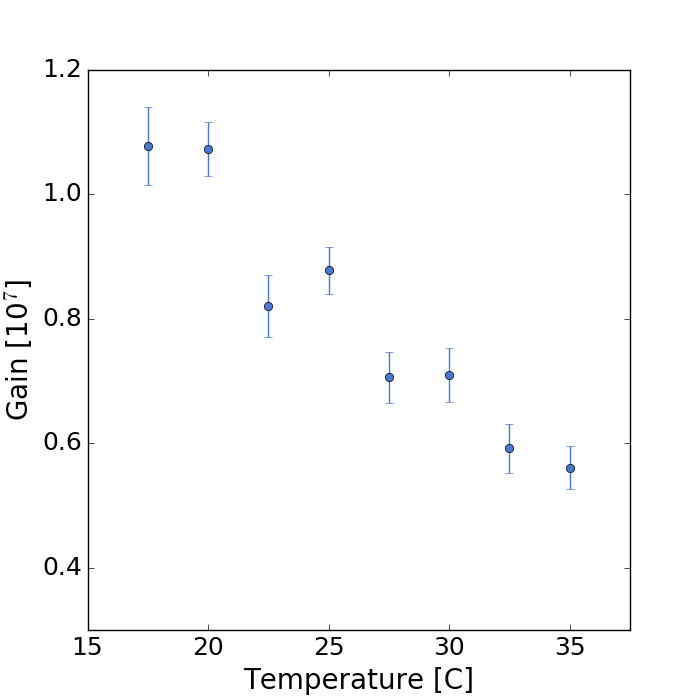}}
\caption{Gain as a function of temperature for one channel (J05).}
\label{fig:gain_temp}
\end{figure}

\begin{figure}[h!]
\centering{
\includegraphics[scale=0.45]{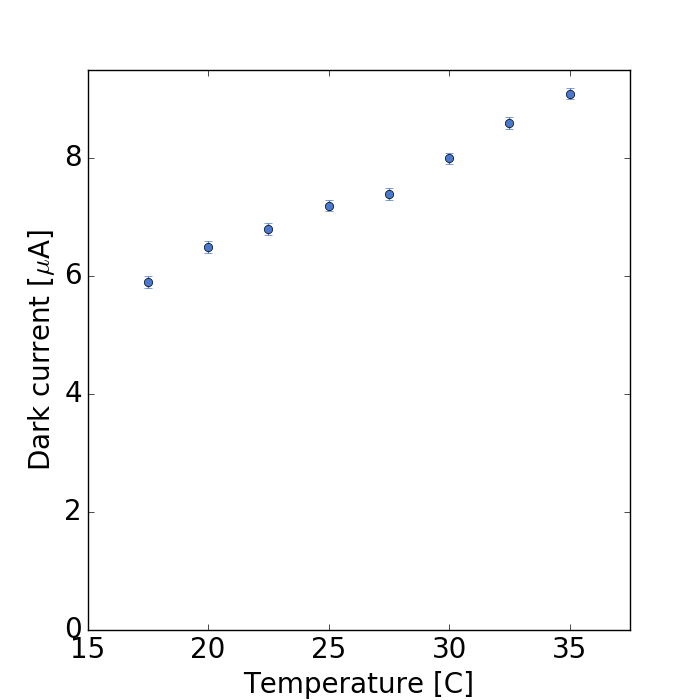}}
\caption{Current drawn by the ANL MCP as a function of temperature.}
\label{fig:dark_current}
\end{figure}

It is important to understand the behaviour of the MCP as a function of the
ambient temperature. This was performed by using a light-tight
climate chamber to measure gain variation with temperature. At each point in the
scan, the temperature was allowed to settle for approximately twenty minutes. The gain
of a single channel was measured at 2580 V and is shown in Figure~\ref{fig:gain_temp}.
From this a 3\% reduction in gain per $^\circ$C between
15$^\circ$C and 35$^\circ$C is observed.

Figure~\ref{fig:dark_current} shows the variation of the dark current drawn by the 
ANL-MCP as a function of the temperature. It shows a linear increase
between 15$^\circ$C and 35$^\circ$C, dominated by the leakage current.

\section{Future plans}

``Tube 32" is an early prototype MCP from ANL in which the
voltage distribution within the device depends completely on the MCP and
spacer resistances, defined by the ALD, but not under control during the
tube processing. ANL have loaned us a next generation device 
where the stack voltages can be independently controlled, potentially
allowing for better tuned performance. This device is now under test.

In addition, we plan to perform further magnetic field tests,
including cross-talk studies in field. Importantly, with a new
laser system and oscilloscope we will perform measurements of the time
resolution.

\section{Summary}

These proceedings have described the testing and characterisation of a
prototype ANL MCP-based photosensors using the test infrastructure
at the University of Edinburgh. The gain of the device was tested as
a function of the high voltage, showing typical gains between $10^6$ and
$10^7$ depending on the channel. The device was shown to be 
insensitive to small (up to 30 G) transverse and longitudinal magnetic
fields. A large decrease in gain was observed as a function of temperature.
Despite the device being a prototype, it is important
to feed this information back to the manufacturer and the study
repeated with next generation devices.
Potential applications of these devices were described,
both for future particle collider and neutrino experiments, which could 
greatly benefit from the excellent timing and spatial resolutions offered by 
MCP photodetectors.

\section{Acknowledgements}

The authors would like to thank Robert Wagner, Jingbo Wang, Junqi Xie and Lei Xia
from Argonne National Laboratory
for the loan of the prototype MCP used in this study and for advice given. GC
acknowledges the support of the Science and Technology Facilities Council grant
ST/K004646/1.

\bibliography{mybibfile}

\end{document}